\documentstyle[12pt,epsfig,amssymb]{article}
\def\Journal#1#2#3#4{{#1} {\bf #2}, #3 (#4)}

\def\NP{{{\em Nucl. Phys.} B}}
\def\PL{{\em Phys. Lett.}  B}

\newcounter{hran}
\def\re#1{(\ref{#1})}
\def\beq{\begin{equation}}
\def\eeq{\end{equation}}
\def\beeq{\begin{eqnarray}}
\def\beeqn{\begin{eqnarray*}}
\def\eeeq{\end{eqnarray}}
\def\eeeqn{\end{eqnarray*}}
\def\bit{\begin{itemize}}
\def\eit{\end{itemize}}
\def\ben{\begin{enumerate}}
\def\een{\end{enumerate}}
\def\nome#1{{\label{#1}}}

\def\bmini{\setcounter{hran}{\value{equation}}
\refstepcounter{hran} \setcounter{equation}{0}
\renewcommand{\theequation}{\thehran\alph{equation}}
              \begin{eqnarray}  }
\def\bminiG#1{
          \setcounter{hran}{\value{equation}}
          \refstepcounter{hran}
          \setcounter{equation}{-1}
          \renewcommand{\theequation}{\thehran\alph{equation}}
          \refstepcounter{equation}
    \label{#1}
          \begin{eqnarray}          }
\def\emini{\end{eqnarray}\setcounter{equation}{\value{hran}}
\renewcommand{\theequation}{\arabic{equation}}}
\newskip\humongous \humongous=0pt plus 1000pt minus 1000pt
  \newif\ifdtup

\newif\ifdtup

\def\a{\alpha}

\def\g{\gamma}                  \def\G{\Gamma}
\def\de{\delta}                 \def\D{\Delta}

\def\l{\lambda}                 \def\L{\Lambda}
\def\m{\mu}

\def\s{\sigma}                  \def\S{\Sigma}
\def\th{\theta}

\newcommand{\DD}{{\cal D}}
\renewcommand{\SS}{{\cal S}}
\newcommand{\OO}{{\cal O}}
\newcommand{\WW}{{\cal W}}
\def\d4#1{\frac {d^4 {#1} }{(2\pi)^4}}

\newcommand{\cint}{\int d^4x\; d^2\th\;}

\newcommand{\zint}{\int_{z}}
\newcommand{\spint}{\int_{p}}
\newcommand{\spqint}{\int_{pq}}

\newcommand{\lp}{\left(}
\newcommand{\rp}{\right)}

\newcommand{\lgr}{\left\{}
\newcommand{\rgr}{\right\}}

\def\tr{\,\mbox{Tr}\,}
\def\frac#1#2{ {{#1} \over {#2} }}
\def\half{\mbox{\small $\frac{1}{2}$}}
\def\p{\partial}

\def\ie{\hbox{\it i.e.}{ }}

\newcommand{\ad}{{\dot\a}}    
\newcommand{\bD}{{\bar{D}}}
\newcommand{\bJ}{{\bar{J}}}

\def\cp{c_{+}}
\def\cm{c_{-}}
\def\bcp{\bar c_{+}}
\def\bcm{\bar c_{-}}

\def\bxi{\bar\xi}
\def\bp{\bar p}

\def\bphi{\bar \phi}

\def\gv{\g_{\scriptscriptstyle{V}}}

\def\bt{\bar\th}

\def\se{S_{\mbox{\footnotesize{eff}}}}

\def\si{S_{\mbox{\scriptsize{int}}}}
\def\scl{S_{\mbox{\scriptsize{cl}}}}
\def\sgf{S_{\mbox{\scriptsize{gf}}}}
\def\sfp{S_{\mbox{\scriptsize{FP}}}}

\def\sym{S_{\mbox{\scriptsize{SYM}}}}

\def\bG{\bar\Gamma}
\def\Gr{\G_{\mbox{\footnotesize{rel}}}}

\def\Gir{\G_{\mbox{\footnotesize{irr}}}}
\def\Gi{\G_{\mbox{\scriptsize{int}}}}

\def\De{\D_{\mbox{\footnotesize{eff}}}}

\def\DG{\D_{\G}}

\def\DGi{\D_{\G,\mbox{\footnotesize{irr}}}}

\def\DGh{\hat{\D}_{\G}}

\def\LdL{\L\partial_\L}

\def\dLID#1{ \frac {\L \partial D^{-1}_{\L\L_0}(#1)}{\partial \L}}

\def\UV{$\L_0\to\infty\;$}

\def\K{K_{\L\L_0}}

\def\Kiu{K_{0\L_{0}}}

\def\t,#1{t^{#1}}
\def\f,#1#2#3{f^{#1 #2 #3}} 
\begin{document}
\begin{titlepage}

\begin{flushright}
UPRF-98-14 \\
November 1998
\end{flushright}
\vspace{.4in}
\begin{center}
{\large{\bf 
SUPERSYMMETRIC GAUGE THEORIES 
IN THE EXACT RENORMALIZATION GROUP APPROACH~\footnote{
Talk given at the {\it
Workshop on the Exact Renormalization Group}, Faro (Portugal), 10-12
September 1998.}}}
\bigskip \\ F. Vian
\\
\vspace{\baselineskip}
{\small Universit\`a degli Studi di Parma \\ and\\
I.N.F.N., Gruppo collegato di Parma, 
\\viale delle Scienze, 
43100 Parma, Italy} \\
\mbox{} \\
\vspace{.5in}
{\bf Abstract} \bigskip \end{center} 
\setcounter{page}{0} 
In these notes the exact renormalization group formulation of the
scalar theory is briefly reviewed. This regularization scheme is then
applied to supersymmetric theories. In case of a supersymmetric gauge
theory it is also shown how to recover gauge invariance, broken by 
the introduction of the infrared cutoff.
\end{titlepage}

\section{Introduction}
The aim of this talk is to discuss qualitatively how the
Wilson renormalization group~\cite{P} (RG)  can be
implemented  in supersymmetric gauge theories.

The RG formulation provides
the most physical framework to study general properties of
renormalized quantum field theories and, in particular, to deal with
ultraviolet (UV) divergences.  Furthermore, it is in this context that
effective theories naturally arise. Recently supersymmetric gauge
theories have been focussed and the definition of a low-energy
supersymmetric Wilsonian effective action has become urgent.  In the
RG approach the bare action $\si(\L_0)$ can be viewed as the result of
integrating out all degrees of freedom with frequencies larger than
the UV cutoff $\L_0$ of a more elementary underlying theory. By
further integrating the fields in the path integral with frequencies
larger than some scale $\L<\L_0$, one obtains the so-called Wilsonian
effective action $\se$. Such an operation can be carried out~\cite{BDM} by
multiplying the quadratic part of the classical action by a cutoff
function $\K(p)$ which falls off sufficiently rapid for $p^2$ outside
the region $\L^2<p^2<\L_0^2$. 

Despite the loss of gauge invariance at the scale $\L$, when all
cutoffs are removed the Slavnov-Taylor (ST) identity can be recovered
(at least in perturbation theory) by properly fixing the boundary
conditions of the RG equation.

The way the cutoffs are introduced in the RG formalism proves
particularly suitable in the supersymmetric case.  If the classical
action is written in terms of superfields, the regularization
procedure described above preserves supersymmetry (in components this
corresponds to use the same cutoff function for all fields).~\cite{BV}
As our formulation works in $d=4$, we can exploit the superspace
technique, which is 
unambiguous and simplifies perturbative calculations.

\section{Renormalization group flow}

To start with, we briefly recall how the RG method works in the
scalar case.~\cite{BDM}  According to Wilson one integrates over the
fields with frequencies $\L^2<p^2<\L_0^2$ in the path integral and
obtains
\beq\nome{Z}
Z[j]=N[j;\L] \;\int {\cal D}\phi \,
\exp i\{
\half (\phi\, D^{-1}\phi)_{0\L}+(j\phi)_{0\L}+\se\,[\phi;\L]
\; \}
\,,
\eeq
where $N[j;\L]$ contributes to the quadratic part of $Z[j]$.  The
functional $\se$ is the Wilsonian effective action and is the
generator of the connected amputated cutoff Green functions (except
the tree-level two-point function). It contains a cutoff propagator
$D_{\L\L_0}$, that is to say the free propagator $D(p)$ is multiplied
by a cutoff function $\K$ which is one for $\L^2 < p^2 < \L_0^2$ and
rapidly vanishes outside.  Notice that at $\L=\L_0$ the Wilsonian
effective action coincides with the bare action. The requirement that
$Z[j]$ is independent of the IR cutoff $\L$ can be translated into a
flow equation for $\se$, referred to as the exact RG equation. For our
purposes it is convenient to perform a Legendre transform on $\se$ in
order to obtain the so-called ``cutoff effective action''
$\G[\phi;\L]$, which is the generator of 1PI cutoff vertex functions
and reduces to the physical quantum effective action in the limits
$\L\to 0$ and \UV.  The next step consists in deriving a flow equation
for $\G[\phi;\L]$, which is straightforward once the RG equation for
$\se$ is given.  It reads
\beeq\nome{eveq}&&
\LdL \biggr\{ \G[\phi;\L]
-\half \int \d4p D^{-1}_{\L\L_0} (p)\,\phi(p)\phi(-p) \biggr\}
\\
\nonumber&&\phantom{\LdL \biggr\{ \G}
=- \frac i2\int \d4q
\dLID {q}\, \frac {1}{\G_2(q;\L)}\,
\bG[q,-q;\phi;\L]\, \frac {1}{\G_2(q;\L)}\,.
\eeeq
The auxiliary functional $\bG$, originating from the inversion of the
functional $\frac{\de^2 W}{\de J\de J}$, satisfies a recursive
equation which  gives $\bG$ in terms of the proper vertices of
$\G$. A graphical representation of $\bG$ is given in fig. 1.
\begin{figure}[htbp]
\epsfysize=1.9cm
\begin{center}
\epsfbox{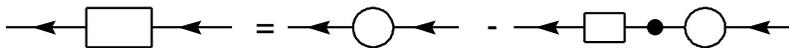}
\end{center}
\caption{
Graphical representation of the auxiliary functional $\bG$. The box
and the blob represent the functionals $\bG$ and $\Gi$, respectively. The
dot indicates a cutoff full propagator.}
\end{figure}
\noindent
Equation \re{eveq}, together with a set of suitable boundary conditions,
can be thought as an alternative definition of the theory which in
principle is non-perturbative. As far as one is concerned with its
perturbative solution, the usual loop expansion is recovered by
solving iteratively \re{eveq}.

\subsection{Boundary conditions}
In order to set the boundary conditions we distinguish between 
relevant  couplings and irrelevant vertices according to their mass dimension.
Relevant couplings have non-negative mass dimension and are 
defined as the value of some vertices and their derivatives 
at a given normalization point. 
For the four-dimensional massless scalar field theory we are considering 
the relevant couplings $\s_i$'s are defined through
$$
\left.\s_1(\L)=\frac {d \G_2(p;\L)}{dp^2}\right|_{p^2=\mu^2}  \,,\;\;\;\;\;\;\;
\s_2(\L)=\G_2(p;\L)\vert_{p^2=0}\,, \;\;\;\;\;\;\;
\s_3(\L)=\G_4(\bp_i;\L)\,,
$$
with $\bp_i$'s  the momenta at the symmetric point
$\bp_i\bp_j=\mu^2(\delta_{ij}-\frac{1}{4})$.
Notice the introduction of the subtraction point $\mu$ to get rid of IR
divergences.

We eventually assume the following boundary conditions:
 
\noindent (i) at the UV scale $\L=\L_0$ the simplest boundary
condition for all the irrelevant vertices is that they 
vanish. As a matter of fact $\G[\phi;\L=\L_0]$ reduces to the bare
action, which must contain only renormalizable interactions  
in order to guarantee perturbative renormalizability;

\noindent (ii)  the relevant couplings are fixed at the physical
point $\L=0$ in terms of the physical couplings, such as the wave 
function normalization, the three-point coupling and the mass.  
Hence the boundary conditions to be imposed  on the relevant couplings 
are 
$$
\s_1(\L=0)=1 \,,\;\;\;\;\;\;
\s_2(\L=0)=0\,,  \;\;\;\;\;\;
\s_3(\L=0)=\l \,.
$$

\section{The Wess-Zumino model}
Now that we have been acquainted with the RG formalism in the scalar
case the extension to a supersymmetric theory is straightforward.
In the Wess-Zumino model the classical Lagrangian reads
$$
S_{cl}= \frac{1}{16}\zint \bphi\; \phi +
\frac{\l}{48}\,\cint  \phi^3 + \mbox{h.c.}\,.
$$
Integration over $z$ means integration over the full superspace
($d^4x\,d^4\th$) and $\phi$ ($\bphi$) is a chiral (anti-chiral)
superfield satisfying $\bD_\ad\phi=0$ ($D^\a\bphi=0$).
Linearity of the supersymmetry
transformation when acting 
on $\phi$ ($\bphi$)~\cite{piguet} is a key ingredient of our formulation.
As we previously did in the scalar case, we follow
Wilson and integrate over the fields with frequencies
$\L^2<p^2<\L_0^2$. By collecting the fields and the sources in 
$\Phi_i=(\phi,\,\bphi)$
and $J_i=(J,\,\bJ)$ respectively,
and introducing the general  cutoff scalar products between 
fields and sources
$$
\half (\Phi,\,\DD^{-1}\Phi)_{\L\L_0}\equiv\frac{1}{16}
\spint\, K^{-1}_{\L\L_0}(p)\,
\bphi(-p,\,\th)\,\phi(p,\,\th)\,,
\,\;\;\;\;
\spint \equiv \int \frac{d^4p}{(2\pi)^4} \,d^2\th \,
d^2\bt
$$
$$
(J,\Phi)_{\L\L_0}\equiv\frac{1}{16}
\spint\, K^{-1}_{\L\L_0}(p) \,
\lgr J(-p,\,\th)\,\frac{D^2}{p^2}\,\phi(p,\,\th)\,
+\bJ(-p,\,\th)\,\frac{\bD^2}{p^2}\,\bphi(p,\,\th)
\rgr
\,
$$
(recall that $\frac{\bD^2 D^2}{16p^2} \phi=\phi$), the generating
functional 
can now be written in
terms of the Wilsonian effective action $\se[\Phi;\L]$ as in \re{Z}.
Then we can derive
the cutoff effective action $\G[\Phi;\L]$ and it turns out it
satisfies an evolution equation very similar to the one we met in the
scalar case, slightly modified by the presence of covariant
derivatives.~\cite{BV} 
The auxiliary functional $\bG$ has
exactly the same  expansion in terms of the vertices of $\G$.

\subsection{Boundary conditions}
We know that at the UV scale $\L=\L_0$ the cutoff effective action
$\G[\Phi;\L=\L_0]$ reduces to the bare action, which contains only
renormalizable supersymmetric interactions since in our formulation
supersymmetry is manifest.~\footnote{The introduction of the cutoff
does not spoil global symmetries as long as they are linearly
realized.}  
On the other hand, the relevant couplings are fixed at the physical
point $\L=0$ in terms of the physical couplings, such as the wave
function normalization, the three-point coupling and the mass.  
Let us consider as an example the massless chiral multiplet two-point
function (\ie the $\phi\bphi$-coefficient of the cutoff effective
action) $\G_{2}(p;\L)=\DD^{-1}\K^{-1}(p)+\S(p;\L)$, which contains the
relevant coupling $Z(\L)=\S(p;\L)
\left. \right|_{p^2=\mu^2}$. The boundary conditions we assume 
are  $Z(\L=0)=0$ and 
$\S_{\mbox{\footnotesize{irr}}}(p;\L)\vert_{\L=\L_0} =\lp\S(p;\L)-Z(\L)\rp
\vert_{\L=\L_0}=0\,.$

\subsection{Perturbative expansion}

The iterative solution of the RG equation for $\G$ automatically
provides a renormalized perturbation theory in $\l$. 
As an example we compute the one-loop two-point function.
The evolution equation for this vertex 
is given by~\cite{BV}
\beeq
&& \spint \bphi (-p,\,\th)\, \L \p_\L \S^{(1)} (p;\, \L)
 \, \phi  (p,\,\th)\,=
\frac i{64}\l^2\,
\spqint
\frac {\K(p+q)\L \p_\L\K(q)}{q^2(p+q)^2}\, \nonumber \\
&&\;\;\;\;\;\;\;\;\;\;\;\;\;\times\,
\bphi (-p,\,\th_1)\, \phi  (p,\,\th_2)\, \de^{4} (\th_1-\th_2)\,
\bD^2 D^2(q,\,\th_2)  \, \de^{4} (\th_1-\th_2)\,.\nonumber
\eeeq
By carrying out some standard $D$-algebra manipulations and
implementing the boundary conditions, the solution of $\S$ at the
physical point $\L=0$ and in the \UV limit is $$
\S^{(1)} (p;\, \L=0) =\frac i{128} \,\l^2\,
\int \frac{d^4 q}{(2\pi)^4}
 \; \lp\frac {1}{q^2\,(p+q)^2}-\left.\frac
{1}{q^2\,(p+q)^2}\right|_{p^2=\mu^2}\rp
\,.
$$
Notice the crucial role of the boundary condition for $Z$, \ie 
$Z^{(1)}(0)=0$,
which naturally provides the necessary subtraction to make the vertex
function $\S$ finite for \UV. 

One immediately realizes that only the vertices with an equal number
of $\phi$ and $\bphi$ are generated at this order.  Anyway 
the coefficients of
$(\phi\bphi)^n$ with $n>1$ are obviously irrelevant and power counting
tells us they are finite, so that no subtraction is needed.

\section{Supersymmetric gauge theories}

When we have to face with (supersymmetric) gauge theories, the
key issue are the boundary conditions, which in addition to ensuring
(perturbative) renormalizability and providing the physical couplings $g(\m)$
have to guarantee symmetry.
In fact the cutoff $\L$ explicitly breaks gauge invariance and we have
to show that at the physical point $\L=0$ (in the \UV limit) the
Slavnov-Taylor identity can be recovered by properly fixing the
boundary conditions, at least in perturbation theory.
 
Let us start with $N=1$ Super Yang-Mills. The classical action reads
$$
\sym  = -\frac{1}{128g^2}\tr\cint \WW^\a \WW_\a + \mbox{h.c.}\,, \;\;\;\;\;\;\;\;
\WW_\a = \bD^2\lp e^{-gV}D_\a e^{gV}\rp\,,
$$
where $V(x,\th)$ is the $N=1$ vector supermultiplet which belongs to
the adjoint representation of the gauge group $G$ ($V=V^a\tau_a$). As
the classical action is gauge invariant, in order to quantize the
theory we have to fix the gauge and choose a
regularization procedure.
Instead of the familiar Wess-Zumino gauge ---in which the
supersymmetry transformation is not linear--- we choose a
supersymmetric gauge fixing. Hence we add  to the action a 
gauge fixing term which is a supersymmetric extension of the
Lorentz gauge and the corresponding Faddeev-Popov term.~\cite{piguet}
The classical action $\scl=\sym +\sgf+\sfp$
is invariant under the usual BRS transformation.

Eventually we take into account matter fields. Matter is described by a
set of chiral superfields $\phi^{\mbox{\tiny{I}}}(x,\th)$ which belong
to some representation $R$ of the gauge group. The BRS action for the
matter fields is
$$
S_{\rm m} = \frac{1}{16} \zint  \bphi\, e^{gV^a T_a}\phi 
$$
plus a possible superpotential $W(\phi)$.

Developing the RG formalism for supersymmetric theories
is straightforward once one replaces the sets of fields and sources
in \re{Z} with 
\beeq 
&&
\Psi_i= (V,\, \cp,\,         \bcm,    \,  \cm,   \, \bcp\,\phi,\,\bphi)
\,,\;\;\;\;\; \g_i=(\gv\,,\g_{\cp}\,,\g_{\bcp}\,,\g_\phi\,,\g_{\bphi})\,,
\nonumber \\ 
&&
J_i=    (J_V, \, \xi_{-}+\bD^2\gv,\, -\bxi_{+},\, -\xi_{+},\, \bxi_{-}-
D^2\gv,\,J,\,\bJ)\,, \nonumber
\eeeq
where  the ghost $\cp$ and the anti-ghost $\cm$ are chiral fields,
like the gauge  parameter.  
The sources $\g_i$ are associated to the BRS variations of
the respective superfields, so that we are able to manage composite
operators.

\subsection{Boundary conditions}

The relevant part of the cutoff effective action 
involves full superspace integrals 
of monomials in the fields, sources and derivatives local in $\th$,
with dimension not larger than two,  
invariant under
Lorentz and global gauge transformations. Due to the dimensionless nature of
the field $V$, $\Gr$  contains infinite terms which can be classified
according to the number of gauge fields.
The couplings $\s_i(\L)$, \ie the coefficients of those monomials, 
can be expressed in terms of the 
cutoff vertices at a given subtraction point, generalizing the
procedure used in  subsect.~3.1 to define the coupling $Z(\L)$.

As usual the  boundary condition we impose on  the 
irrelevant part of the cutoff effective action is $\Gir[\Phi,\g;\L=\L_0]=0$.
The way in which the boundary conditions for the relevant couplings
$\s_i(\L)$ are determined is not straightforward. In the case of a gauge theory
there
are interactions in $\Gr$ which are not present in $\scl$, so
that only some of the relevant couplings are connected to the physical
couplings (such as the wave function normalizations and the
three-vector coupling $g$ at a subtraction point $\mu$). Therefore, in order to fix the boundary conditions for all the
relevant couplings, one needs an additional  fine-tuning procedure which
implements  the gauge symmetry at the physical point.

\subsection{Gauge symmetry}
The gauge symmetry requires that the physical effective action
satisfies the ST identity. One can show that 
for the Wilson effective action $\se$
such identity  can be 
rephrased  as
$$
\SS_J Z[J,\g]=N[J,\g;\L] \int \DD\Psi e^{i\lgr
\half(\Psi,\DD^{-1} \Psi)_{0\L}+(J,\Psi)_{0\L} 
+\se[\Psi;\L]\rgr} \, \De[\Psi,\g;\L]\, ,
$$
where $\SS_J$  is the usual ST operator.
Restoration of symmetry, $\SS_J Z[J,\g]=0$, translates into 
$$
\De[\Psi,\g;\L]=0 \;\;\;\; {\mbox{for any}} \;\;\L\,.
$$ 
From a perturbative point of view, instead of studying $\De$ it is
convenient to introduce its Legendre transform $\DG$, in which
reducible contributions are absent. We refer to literature for the
derivation of $\DG$.~\cite{MT} For our purposes it suffices to observe
that $\DG$ is made up of two pieces, the first being essentially the
standard ST operator applied to $\G(\L)$, the second, $\DGh$, being
$\OO (\hbar)$ and vanishing for $\L \to 0$.  The functional $\DGh$ at
the UV scale is schematically represented in fig. 2.
\begin{figure}[htbp]
\epsfysize=3.5cm
\begin{center}
\epsfbox{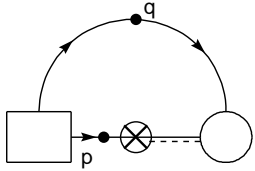}
\end{center}
\caption{
Graphical representation of $\DGh(\L_0)$. The box and 
the circle represent the functionals $\bG$ and
$\G$ respectively. The top line is the cutoff full propagator of the 
field $\Psi$;
the bottom full line represents the field $\Psi$ while the double
line is the corresponding BRS source $\g$.
The cross denotes the product of the two functionals with the
insertion of the cutoff function $\Kiu(p)$.
Integration over the loop momentum is understood.}
\end{figure}
\noindent
What we have to show to recover symmetry is we can set $\DG=0$ in
perturbation theory. Hence we need the evolution equation for such a
functional.~\cite{MT} It turns out the evolution of the vertices of $\DG$ at
the loop ${\ell}$ depends on vertices of $\DG$ itself at lower loop
order, so that if $\DG^{(\ell')}=0$ at any loop order $\ell'<\ell$,
then $\DG^{(\ell)}$ is constant. 
Thus one can analyse $\DG$ at an arbitrary value of $\L$.  
A natural choice 
is  $\L=\L_0$, since at this
scale $\DG$ is local,~\cite{BV} or, more precisely, $\DGi(\L_0)={\cal O}(\frac1{\L_0})$. Once the locality of $\DG(\L)$ is shown, the solvability of the
equation $\DG(\L)=0$ can be proven using cohomological methods.~\cite{piguet}  This is a consequence of the $\L$-independence
of $\DG$ and the solvability of the same equation at $\L=0$, where the
cohomological problem reduces to the standard one.

The requirement that the physical effective action satisfies the ST identity
translates at  $\L=\L_0$ into the fine-tuning equation 
\beq \nome{fintun}
{\cal S}_{\G}\,\G(\L_0) = - \DGh(\L_0)\,,
\eeq
which allows to determine the couplings of the relevant, non-invariant
interactions in $\G(\L_0)$. Once the field normalization and the gauge
coupling are fixed at the physical point $\L=0$, the problem of
assigning the boundary conditions of the RG equation is finally
solved.

However, when the matter representation is such that the matching
conditions for the anomaly cancellation are not fulfilled, we tailored
a simple method to compute the chiral anomaly.~\cite{BV} In our
framework a violation of the ST identity results in the impossibility
of fixing the relevant couplings in $\G(\L_0)$ in such a way that
\re{fintun} is satisfied.  What happens is that part of the
matter contribution to the vertices $\cp$-$V^n$ and $\bcp$-$V^n$ of $\DGh$
yields a cutoff independent result with no match in the l.h.s. of
\re{fintun}, which is precisely the anomaly.~\footnote{It is known 
the anomaly can be expressed as an infinite series in the gauge field $V$.} 

Finally, though we restricted our analysis to the perturbative regime,
the RG formulation is in principle non-perturbative and provides a
natural context in which to clarify the connection between exact
results and those obtained in perturbation theory. In particular, 
the relation between the Novikov-Shifman-Vainshtein-Zakharov $\beta$
function for the gauge coupling  in $N=1$ supersymmetric gauge
theories and the exact 1-loop $\beta$
function for the gauge coupling in a Wilsonian action is presently
under investigation.

\section*{Acknowledgments}
I am grateful to M. Bonini for collaboration on the subject presented
in this talk. I also wish to thank the organizers of the Faro meeting for
providing a stimulating environment and all the participants
for illuminating discussions.

\end{document}